# A Multi-Dimensional approach towards Intrusion Detection System

**Manoj Rameshchandra Thakur***

Computer Science Department,
VJTI, Mumbai, India
manoj.thakur66@gmail.com

*Corresponding author

**Sugata Sanyal**

School of Technology and Computer Science,
Tata Institute of Fundamental Research, Mumbai,
India
sanyals@gmail.com

## ABSTRACT
In this paper, we suggest a multi-dimensional approach towards intrusion detection. Network and system usage parameters like source and destination IP addresses; source and destination ports; incoming and outgoing network traffic data rate and number of CPU cycles per request are divided into multiple dimensions. Rather than analyzing raw bytes of data corresponding to the values of the network parameters, a mature function is inferred during the training phase for each dimension. This mature function takes a dimension value as an input and returns a value that represents the level of abnormality in the system usage with respect to that dimension. This mature function is referred to as *Individual Anomaly Indicator*. *Individual Anomaly Indicators* recorded for each of the dimensions are then used to generate a *Global Anomaly Indicator*, a function with n variables (n is the number of dimensions) that provides the *Global Anomaly Factor,* an indicator of anomaly in the system usage based on all the dimensions considered together. The *Global Anomaly Indicator* inferred during the training phase is then used to detect anomaly in the network traffic during the detection phase. Network traffic data encountered during the detection phase is fed back to the system to improve the maturity of the *Individual Anomaly Indicators* and hence the *Global Anomaly Indicator.*

## 1. INTRODUCTION
### 1.1 Background
Security breach in the recent past has been a major cause of data loss, illegal data access and malicious usage of computing resources which are part of enterprise networks. There has been a tremendous increase in not only the number of instances of security breaches but also the intensity of the impact of these breaches. In 2000 there were 5 incidences of security breaches in which approximately 315000 data records were compromised. Whereas in the year 2008 the number of incidences of security breaches increased to 275-280 in which approximately 170000000 data records were compromised [2]. With the introduction of the cloud infrastructure for storing and processing vital information, security and authorization have become two major concerns. The advent of social networking websites has caused a major rise in the internet traffic. Moreover, with the advancement of the mobile platform as a potential means of accessing several services hosted on the web the internet traffic has increased exponentially in recent times[8][9]. With such high volume of internet traffic generated daily it is extremely difficult to predict exactly the occurrence of an intrusion. This is primarily because the malicious traffic forms a very small portion of the internet traffic and the pattern it follows varies based on the kind of attack used to compromise the target system. It is, however, important to note that in most of the illegal and malicious intrusions, the network traffic values and system usage data roughly follow a pattern with respect to certain features.

Detection of abnormality in the system in such a scenario is feasible if each of the network traffic and system usage features are analyzed separately as individual 'dimensions' and then the combined effect of each of these dimensions is considered. For example, in case of a UDP flood attack if the 'incoming network traffic flow' and the 'protocol' of the incoming packets are analyzed as two separate dimensions then it can be seen that a UDP flood attack is characterized by a combined effect of high 'incoming network traffic flow' which corresponds to the incoming UDP (protocol) packets [16]. The suggested approach thus tries to analyze the network traffic and system usage features as separate dimensions using Soft Computing and Machine learning techniques and then deduces the combined effect of each of the dimensions on the abnormality in usage of the system. The analysis of each of the dimensions includes generation of regression functions for each of the dimensions during the training phase. These regression functions are then applied in the detection phase to detect abnormality in system usage. The regression functions are generated using supervised learning techniques. Data encountered during the detection phase is fed back to the system to improve the regression functions.

## 1.2 Related Work

Due to the lack of exactness and inconsistency in the network traffic patterns a number of approaches towards intrusion detection based on 'Soft Computing' [11] have been proposed [12]. These works attempt to develop inexact and approximate solutions to the computationally-hard task of detecting abnormal patterns corresponding to an intrusion. [4] presents a Soft Computing based approach towards intrusion detection using a fuzzy rule based system. [13] tries to detect abnormality in electromagnetic signals in a complex electromagnetic environment which is similar to the vast and complicated World Wide Web. [15] suggests an approach based on machine learning techniques for intrusion detection. [37] applies a combination of protocol analysis and pattern matching approach for intrusion detection. [18] proposes an approach towards intrusion detection by analyzing the system activity for similarity with the normal flow of system activities using classification trees. A number of works related to intrusion detection have been oriented around the similarity between the human immune system and an Intrusion Detection System. [20] is one of the first lightweight intrusion detection systems based on AIS (Artificial Immune System) [19]. Most of the AIS models suggested have been based on the following algorithms; negative selection algorithm, artificial immune network, clonal selection algorithm, danger theory inspired algorithms and dendritic cell algorithms [32]. [21] attempts to develop an intrusion detection system based on the emerging 'Danger Theory' [30]. [1] tries to draw an analogy between the human immune system and the intrusion detection system. An important feature of this work is that it attempts to evolve the Primary Immune Responses to a Secondary Immune Response using genetic operators like selection, cloning, crossover and mutation. [10] suggests a proactive detection and prevention technique for intrusions in a Mobile Ad hoc Networks (MANET). [38] describes a genetic classifier-based intrusion detection system.

The rest of the paper is structured as follows: Section 2 introduces the concept of multi-dimensional approach and its relevance in Intrusion Detection. Sections 3,4,5 explain the training, detection and feedback phase of the suggested approach. Section 6 explains the propagation of information related to intrusion detection, based on the suggested approach, across the nodes of a network followed by the Conclusion in section 7.

## 2. THE MULTI-DIMENSIONAL APPROACH

Multi-dimensional analysis of data is an important part of Business Intelligence (BI). Multidimensional analysis of business critical data reveals Key Performance Indicators (KPI) that are important for organizations to understand various trends and patterns in business [27]. Analysis of data with respect to multiple dimensions yields significant advantages not only in businesses but also in various fields of study such as engineering, medicine, geology etc. [27][28]. The suggested approach attempts to incorporate the multi-dimensional view of data to analyze network traffic and system usage features for anomaly detection in enterprise networks and distributed infrastructures. The network traffic pattern that determines the anomaly in the usage of the system is characterized by multiple features. The suggested approach attempts to analyze network traffic by considering these features as different dimensions and developing functions for each of these dimensions. The output of the functions for each of the dimensions is then used to generate a global function to calculate the anomaly in the usage of the system as a whole.

## 2.1 Dimension Selection

An important aspect of the suggested multi-dimensional approach is the selection of the network, system usage features that should be considered as different dimensions. The process of selection of appropriate dimensions is similar to the process of feature selection in various machine learning techniques [33]. We use the network and system usage features for which the training data represents a particular pattern. Figure 1 shows two graphs of network parameter $x$ v/s the abnormality $y$ (could be a value in a particular range). In the first graph, the data roughly reveals a linear relationship between abnormality $y$ and variable $x_1$ of the form:

$$ax + by = c$$

Due to this linear relation, parameter $x_1$ can be considered as a potential dimension. The parameter $x_2$, however cannot be considered as a potential dimension. This is because of the lack of relationship between the parameter $x_2$ and abnormality $y$ as shown in the second graph. It is important to note that the graphs shown in Figure 1 are used merely to explain the process of selection of dimensions and are not based on any real time network traffic data.

**Figure 1 Selection of appropriate network parameters as dimensions**

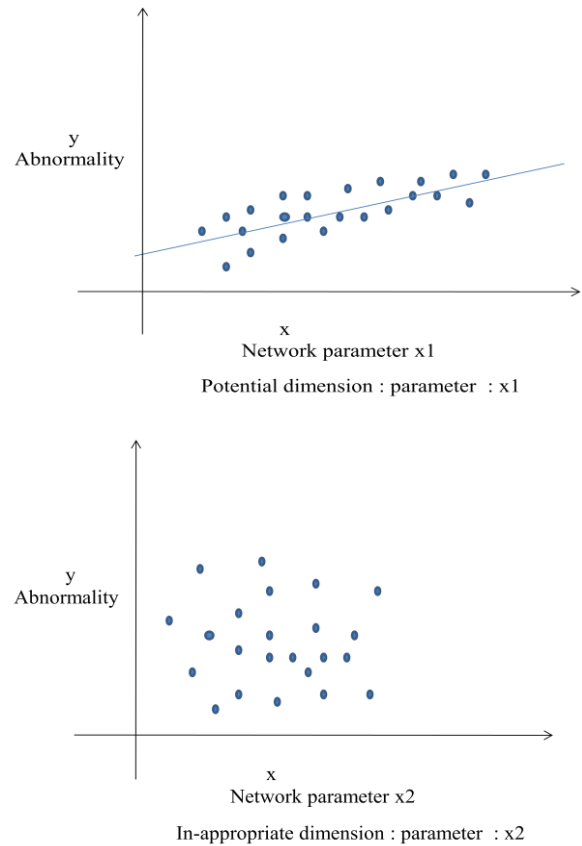

We use Principal Component Analysis [34] to identify features that account for the variability in the training dataset and hence potential patterns that describe the training data set. The covariance matrix for the data set consisting of $n$ variables is calculated. Corresponding to this covariance matrix, $n$ different eigenvectors and their corresponding eigenvalues are recorded [36]. Eigenvectors corresponding to $p$ highest eigenvalues are chosen as they account for the maximum variability of the dataset. Using these $p$ eigenvectors a feature vector is calculated such that each column of the feature vector corresponds to one of the eigenvectors. The transformed dataset is derived by projecting the original data along the $p$ new dimensions [34]. An important advantage of using Principal Component Analysis is that it enables us to efficiently perform analysis on the dataset for only those features that characterize the dataset. [35] attempts to use Principal Component Analysis to reduce the dimensionality of the network traffic data and perform analysis on the transformed data. In our suggested approach, we attempt to not only reduce the dimensionality of the network traffic data using Principal Component Analysis but also perform a two level analysis on the transformed network traffic and system usage data.

## 2.2 Terminologies and Data Representation

Each of the network traffic and system usage feature is considered as a different dimension. Some important properties that characterize the network traffic and system usage are destination and source IP addresses; destination and source ports; incoming and outgoing data rate and the average number of CPU cycles to process each request. The function generated for each of the features (dimensions) are referred to as *Individual Anomaly Indicator* $f_i(x)$, where $f_i$ represents the *Individual Anomaly Indicator* for the $i^{th}$ dimension $d_i$. Each *Individual Anomaly Indicator* returns an *Individual Anomaly Factor* $af_i$ which represents the amount of anomaly in the usage of the system with respect to the dimension $d_i$. The *Individual Anomaly Factors* $af_i$ calculated during the training phase are used to generate the *Global Anomaly Indicator F* that provides the *Global Anomaly Factor AF* for anomaly in usage of the system as a whole. It must be noted that in case of network traffic and system usage parameters like data rate and CPU cycles per request, we consider the per second cumulative or average value as the readings for the training dataset. As explained above the relation between the above terms is as follows:

$$f_i(x_i) = af_i$$

where $x_i$ is the $i^{th}$ value for the dimension $d_i$.

$$F(f_1(x_1), f_2(x_2), f_3(x_3) \dots f_n(x_n))$$
$$= F(af_1, af_2, af_3 \dots af_n) = AF$$

Where $af_1, af_2, af_3 \dots af_n$, represent the *Individual Anomaly Factors* for each of the n dimensions and *AF* is the *Global Anomaly Factor*.

In many scenarios there exists a relationship between multiple network traffic parameters itself. For example if the number of active users in a particular region or an IP subnet $SB_1$ is high then the amount of incoming traffic would be high when the incoming requests are from $SB_1$. Such high incoming traffic originating from another subnet say $SB_2$ which has less number of active users as compared to $SB_1$, is considered abnormal. Thus in such scenarios a feature could contain a combination of network traffic parameters. Thus *Individual Anomaly Indicator* in such a scenario can be represented as:

$$f_i(x_i, y_i, z_i \dots) = af_i$$

Where $x_i$, $y_i$ and $z_i$ represent the network traffic parameters, which are considered together for the dimension $d_i$. Thus $f_i$ in this case is considered as a function with m variables where m represents the number of network traffic parameters considered for the $i^{th}$ dimension.

Each phase involved in the suggested approach is explained below using data obtained through simulations. It must be noted that the simulations are carried out on a limited scale and hence do not represent a practical production scenario. The primary motive has been to manifest the *Individual and Global Anomaly Indicator* generation and the use of these indicators to detect anomaly in the system usage.

## 3. TRAINING PHASE

In this phase, values for each of the features and the anomaly in system usage are recorded and various statistical methods are used to develop a mature Individual Anomaly Indicator. The Individual Anomaly Indicator may vary during the training phase as more data is fed to the system and slight enhancements are made to the *Individual Anomaly Indicator* during the feedback phase from data encountered during the detection phase. The training data set is generated using the readings from a host for the incoming network traffic consisting of UDP packets in bytes/sec. The host was targeted intermittently with UDP flood attack during the time interval when the readings were taken. Thus if the Abnormality Indicator is 1 it represents the presence of an ongoing UDP flood attack. The format of the data set is shown in Table 1.The *Individual Anomaly Indicator* generation and analysis of the same, discussed in the succeeding sections will be based on this training data set.

**Table 1 Subset of the readings for the incoming data rate [in bytes/sec]**

| Input Data rate, *udp* (bytes/sec) | Abnormality Indicator, $a_i$ |
|---|---|
| 1249 | 0 |
| 1067 | 0 |
| 64703 | 1 |
| 53129 | 1 |
| 72637 | 1 |
| 15901 | 0 |

For the training dataset we performed the Principal Component Analysis [34] to identify the suitable features. The eigenvalues, percent variability and cumulative percent variability corresponding to the two features is shown in Table 2.

**Table 2 Eigenvalues, variability for the features.**

|  | F1 | F2 |
|---|---|---|
| Eigenvalue | 1.595 | 0.405 |
| Variability (%) | 79.766 | 20.234 |
| Cumulative % | 79.766 | 100.000 |

**Figure 2 Scree plot indicating cumulative % variability for features F1 and F2.**

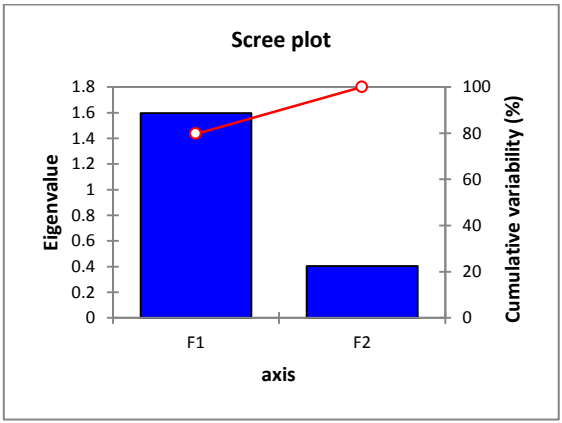

From Table 2 and the scree plot (a plot of features along the X-axis and the corresponding eigenvalues and cumulative variability along the Y-axis) shown in Figure 2, it can be seen that *F1* corresponds to a higher eigenvalue and hence accounts for higher variability in the data as compared to *F2*. Thus we choose *F1* as the dimension for analyzing the anomaly in the system. A subset of the transformed training data using *F1* is as follows:

**Table 3 Subset of the transformed training dataset, projected along the feature F1**

| Observation | F1 |
|---|---|
| 0 | -0.603 |
| 0 | -0.606 |
| 0 | -0.603 |
| 0 | -0.606 |
| 0 | -0.606 |
| 1 | 1.383 |
| 1 | 1.385 |
| 1 | 1.383 |

It must be noted that transforming the training data by projecting it along a particular feature or a set of features is performed as suggested in [34]. We would be using the transformed training data for further analysis. Since the training set consists of labeled data, we use supervised learning technique to determine the *Individual Anomaly Indicator*. Depending on the number of features selected based on the Principal Component Analysis, the techniques used for inferring the *Individual Anomaly Indicator* may vary. For cases when the number of features selected is high, 'Decision Tree' technique [25] can be used to infer the *Individual Anomaly Indicator* by iterating through the training data set. For cases when the number of features selected is fewer, statistical and sequential covering techniques [29] can be employed apart from the 'Decision Tree' technique. For the above training data consisting of a single feature we demonstrate the calculation of the *Individual Anomaly Indicator* using the following two techniques:

1) Decision Tree

2) Sequential Learning

The rationale behind considering these two techniques is as follows:

1) Values for network traffic and system usage parameters and hence the derived features represent continuous and sequential labeled data, sequential learning techniques help to detect the label corresponding to an unknown observation based on the sequence encountered till the unknown observation [26].
2) Abnormality with respect to the features derived using Principal Component Analysis can be effectively expressed using *if then* rules. These *if then* rules can be effectively developed using a decision tree.
3) Regression trees not only help to infer *if then* rules but also help in determining potential dimensions that can be used for detecting the abnormality in system usage [24].

## 3.1 *Individual Anomaly Indicator* Generation using Decision Tree

For the training data set mentioned above we use the Rpart package of the R-project [24] on the transformed dataset to generate a decision tree, using regression, which fits the training data and represents the relation between the selected feature and the system abnormality. The Decision Tree generation using the R-part package is as follows [22][23]:

```
> require(rpart)
> udp.df = read.csv("D:\\testData.csv")
> udp.rpart1 = rpart(formula = Observation ~
F1, data = udp.df, method = "class")
> summary(udp.rpart1)
---------------------------------------------
Call:
rpart(formula = Observation ~ F1, data =
udp.df, method = "class")
n= 1586
    CP nsplit rel error xerror      xstd
1 1.00    0       1       1  0.04780099
2 0.01    1       0       0  0.00000000
Node number 1: 1586 observations,
complexity param=1
predicted class=0  expected loss=0.2162673
   class counts:  1243    343
   probabilities: 0.784 0.216
left son=2 (1243 obs) right son=3 (343 obs)
Primary splits:
     F1 < 0.6005 to the left,
improve=537.6406, (0 missing)

Node number 2: 1243 observations
predicted class=0  expected loss=0
   class counts:  1243     0
   probabilities: 1.000 0.000
Node number 3: 343 observations
predicted class=1  expected loss=0
   class counts:     0   343
   probabilities: 0.000 1.000
```

**Figure 3 Decision Tree based on regression of the training data set**

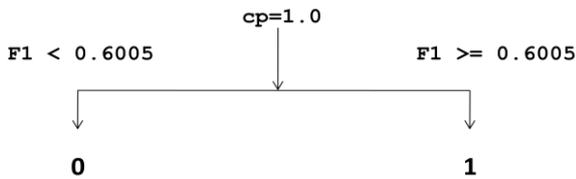

In the Decision Tree generation process the reading were stored in a .csv file named *testData.csv* with two columns *Observation* and *F1*. Each entry in *testData.csv* represents the training data transformed based on the feature F1.The total elements in the data set were 1586. In the summary of the decision tree generated above, complexity parameter (cp) represents the cost complexity factor representing the amount of information required to classify elements into their respective classes [22][23][24][25]. The lower the value of cp the greater is the amount of fitness of the classification tree for the training data set. The classification represented in Figure 3 is fit for the training data set since after the first split the value of cp decreases from `1.00` to `0.01`. Moreover it can be seen that the probabilities for the abnormality values that each of the children in the tree represent, is high indicating a high amount of fitness. The *expected loss* values for each of the children are calculated based on the definition of loss function mentioned in section 3.3.

### 3.2 *Individual Anomaly Indicator generation using Sequential learning*

Sequential learning techniques attempt to infer rules by iterating through the training data set and refining the rule set in each iteration by considering only the subset of the training data that are not described by any of the rules currently existing in the rule set [29].The pseudo code below applies sequential learning technique to the training data set D modifying the rule that describes the training data in each iteration. It is important to note that modifying the rule implies incorporating new rules to the already existing rule R inferred till the current iteration. Modification to a rule could, for example, involve changing the range of the incoming data rate for which abnormality in system usage is reported.

```
function Rule trainSystem( D , Rule , Satisfied-List, UnSatisfied-List ){
            Sort(D);
            for(int I = 0 ; I < D.length ; i++ ) {
            if(Rule == ϕ) {
                    Rule = Generate rule based on only D[i];
                    Satisfied-List.add(D[i]);
            }
            else {
                    if( D[i] satisfies Rule) {
                            Satisfied-List.add(D[i]);
                    }
                    else {
                            UnSatisfied-List.add(D[i]);
                            if(UnSatisfied-List.length >= Satisfied-List.length)
                            {
                                    Rule = adjust Rule for D[i];
                                    Remove elements from UnSatisfied-List which satisfy the adjusted
                                    Rule;
                            }
                    }
            }
    }

    if(UnSatisfied-List.length > Satisfied-List.length * 0.2){
            return trainSystem( UnSatisfied-List, Rule, {}, {} )
    }
    else {
            return Rule;
    }
 }
```

For the above algorithm, at the end of every iteration, a check is made as to whether the number of unsatisfied entries is greater than 20 percent of the number of satisfied entries. This ensures that the flow exits and returns the final rule R only if 80% or higher number of elements in the data set satisfies the returned rule R. Analysis of a subset of the training data set based on the above algorithm is shown in Table 4.

**Table 4 Analysis of a subset of the training data set based on sequential learning**

| Observation | F1 | Rule | Is satisfied? | Satisfied list | Unsatisfied list |
|---|---|---|---|---|---|
| 0 | -0.603 | if(f1 <= -0.603) return 0; else return 1; | Y | {-0.603} | |
| 0 | -0.606 | if(f1 <= -0.606) return 0; else return1; | N | {-0.603} | {-0.606} |
| 1 | 1.383 | | Y | {-0.603,1.383,-0.606} | |
| 0 | -0.333 | | Y | {-0.603,1.383,-0.606,-0.333} | |
| 1 | 1.383 | | Y | {-0.603,1.383,-0.606,-0.333,1.383} | |
| 0 | -0.606 | | Y | {-0.603,1.383,-0.606,-0.333,1.383,-.606} | |
| 0 | -0.603 | | Y | {-0.603,1.383,-0.606,-0.333,1.383,-.606,-0.603} | |

Since after the first iteration the satisfied list count is 7 and the unsatisfied list count is 0 (after adjusting the rule in the second step) we return the rule '*if (f1 <= -0.606) return 0; else return 1;*' as it is within our permissible range of 20%.

Based on the above algorithm the function generated for the whole training data set is as follows:

*f(x) = 0 for x <=0.603*

*= 1 for x >= 0.603*

It must be noted that the algorithm mentioned above is specific to the feature derived using the Principal Component Analysis for the training data set. The anomaly indicator i.e. the label for the dataset is a Boolean value where 1 indicates an anomaly in the system usage and 0 indicates otherwise. The method to calculate the *Individual Anomaly Indicator* may however vary based on the number and type of feature selected. Moreover the anomaly indicator for the training data set could also be a value, lying in a particular range, indicating the extent of anomaly in the system.

For calculating the *Global Anomaly Indicator* the training data set consisting of the values for each of the features (dimensions) is considered. *Individual Anomaly Factors* calculated are then used to infer the *Global Anomaly Indicator* using supervised learning techniques [3]. The process of generation of *Individual Anomaly Indicator* is similar to the Soft Computing based analysis suggested in [4], the evolution of Secondary Immune Response as suggested in [1] is analogous to the mature *Individual Anomaly Indicator* which given a value for a particular feature (dimension) returns the *Individual Anomaly Factor* indicating the abnormality in the system with respect to that feature. However our suggested approach differs as the supervised learning techniques are applied at two levels: firstly at the feature level and then at the global level considering all the features together.

### 3.3 Individual Anomaly Indicator fitness

Corresponding to the same training data set, there can be multiple functions possible that can potentially be used as an *Individual Anomaly Indicator*. Thus to determine the function that best fits the training data we use the concept of *Loss Function, Risk* and *Empirical risk minimization* [3][5]. Consider a function *g* that is chosen as a potential *Individual Anomaly Indicator*. For this function we calculate the *loss function L* defined as:

$$L : Y \times Y \to \mathbb{R}^{\geq 0}$$

For each element {$x_i$, $y_i$} in the dataset the *loss function* indicates the difference between the expected output y` calculated using *g* and the actual output y. This definition of the *loss function* also applies to training set consisting of multiple network traffic parameters which are considered as a single feature (dimension). The empirical risk *R* is then defined as follows [3][5][14]:

$$R_{emp}(g) = \frac{1}{N} \sum_i L(y_i, g(x_i))$$

In the above definition N is the number of elements in the training data set. Thus based on the above definitions *Remp* indicates the fitness of the selected function *g*. The lower the value of *Remp* the greater is the amount of fitness of *g*. For a given dimension the function *g* with the minimum *Remp* is selected as the *Individual Anomaly Indicator* for that dimension. For the *Individual Anomaly Indicator* calculated in the previous section using the Decision Tree techniques the value for the expected loss for each of the left and right children is 0, moreover the expected loss before the split is `0.2162673` indicating a high amount of fitness of the inferred *Individual Anomaly Indicator*.

Once the *Individual Anomaly Indicators* for each of the dimensions is calculated, each of the *Individual Anomaly Indicators* is then used to generate the *Global Anomaly Indicator*. The *Global Anomaly Indicator,* like *Individual Anomaly Indicator* may vary during the training phase. It must be noted that considering network traffic features as different dimensions provides a significant advantage with respect to generating the *Individual Anomaly Indicator.* Even if a new dimension is introduced after the *Individual Anomaly Indicators* are calculated for existing dimensions, the existing *Individual Anomaly Indicators* are not affected and are not required to be generated again. Thus the only additional computations that need to be performed are:

1) Generating the *Individual Anomaly Indicator* for the new dimension introduced.
2) Modifying the *Global Anomaly Indicator* as a result of the introduction of the new *Individual Anomaly Indicator.*

## 4. DETECTION PHASE

In this phase the *Individual Anomaly Indicator* generated during the training phase for each of the dimensions is applied to live network traffic data to get the *Individual Anomaly Factors*. The *Individual Anomaly Factors* calculated are then applied to the *Global Anomaly Indicator* to calculate the anomaly in the system usage. It must be noted that the live network traffic data encountered is first transformed using the feature matrix before any further analysis.

## 5. FEEDBACK PHASE

The suggested approach provides the advantage of not only evolving and improving the *Global and Individual Anomaly Indicators* during the training phase but also by using live production data during the detection phase as a feedback. The feedback not only helps to improve the *Global and Individual*

*Anomaly Indicators* but also helps to identify potential dimensions that can provide information related to the abnormality in system usage. Along with the growing volume of the internet traffic there has been an evolution of various patterns that the internet traffic follows. As a result it is possible that potential dimensions and corresponding *Individual Anomaly Indicators* may not be detected during the training phase. It is in this phase that such possible dimensions surface, which can be used to detect anomalies more accurately.

# 6. PROPOGATION OF ANOMALY INDICATORS ACROSS A NETWORK

In a distributed environment where multiple nodes of a network expose the same set of services or are part of the same distributed infrastructure, the nature of network traffic and system usage pattern is almost similar for each of the nodes of the network. Anomaly detection in such a distributed environment can be achieved by co-operative intelligent agents residing on each of the network nodes, analyzing the traffic and system usage based on the suggested multi-dimensional approach **[31]**. In such a scenario, it is important to maintain consistency as far as the *Global and Individual Anomaly Indicators* inferred at each of the nodes is concerned. Moreover it is also important to propagate information related to malicious nodes in the network which are inferred based on the suggested approach. Consider a scenario where source IP address of incoming packets and incoming data rate corresponding to these set of IP addresses are analyzed to detect a potential DDoS (Distributed Denial of Service Attack), in this scenario the set of malicious nodes detected using the suggested approach at each node can then be propagated to other nodes of the network using the reputation based approach suggested in [6], wherein each node will propagate the malicious node information to only it's immediate neighbors**.** The nodes on receiving this information can take appropriate actions like updating their malicious nodes list, propagating the warning received to their immediate neighbors and updating their *Individual and Global Anomaly Indicator***.** Security of the data (*Individual and Global Anomaly Indicator* and malicious node list) exchanged between the nodes of the network can be ensured by transforming and fragmenting the data packets before transmitting them as suggested in [7]. [17] suggests a similar approach towards secure data transfer over a network using the concept of *jigsaw puzzle*. Each of the approaches [7] [17] ensure that no intermediate node or an unauthorized party can access the data being transferred to its completeness. The data fragments at each of the receiving nodes are then combined and required actions are performed based on the received combined data**.**

# 7. CONCLUSION

In conclusion, the suggested approach is instrumental in detecting abnormalities in the system. A significant improvement and efficiency is achieved using the suggested approach. The concept of dimension facilitates the analysis of the system features based on only the relevant and meaningful network traffic and system usage features. The *Individual and Global Anomaly Indicators* supplant the need to store huge volumes of training data after the training phase. All that needs to be stored during the detection phase are the mature functions developed during the training phase. The approach also facilitates flexibility during the training phase as mature functions for each of the dimensions can be developed independently. Finally the suggested approach also addresses the issues related to application of the multi-dimensional approach in a distributed environment.

## ACKNOWLEDGEMENT
Authors would like to thank Dipankar Dasgupta and Chandrakant Sakharwade for their support and advice.